\documentclass{amsart}
\usepackage{amssymb}

\title{Uniqueness of an $E_8$ model of elementary particles} 
\author{Robert Arnott Wilson}
\date{First draft: 24th July 2024. This version: 31st August 2024.}

\address{Queen Mary University of London}
\email{r.a.wilson@qmul.ac.uk}

\newcommand{\RR}{\mathbb R}
\newcommand{\CC}{\mathbb C}
\newcommand{\HH}{\mathbb H}
\newcommand{\OO}{\mathbb O}
\newcommand{\rep}{\mathbf}

\begin{document}
\begin{abstract}
There are many ways to embed the Lie groups of the Standard Model of Particle Physics in a Lie group of type $E_8$,
but so far there is no convincing demonstration that the finite symmetries (and asymmetries) of weak hypercharge, three generations of electrons,
three quarks in a proton, and photon polarisation can also be embedded correctly. I show that there is a unique way to embed these
finite symmetries consistently, and that the gauge groups of the Standard Model are then uniquely determined. The model is automatically
chiral, and the generation symmetry acts as a rotation in a real 2-space, so that the spinors for three generations have only twice as many
degrees of freedom in total as
the spinors for a single generation. 
In fact, two distinct generation symmetries arise from the restriction to the Standard Model, related by the CKM and/or PMNS matrices.
It therefore appears that these two matrices are not independent. 
I further speculate on the implications for quantum gravity.
\end{abstract}
\maketitle

\section{Introduction}
The largest exceptional Lie group, $E_8$, has long held a fascination for physicists, and many attempts have been made to construct 
a `unified' theory of fundamental physics out of it, or within it or its Lie algebra. 
These range from traditional Grand Unified Theories (GUTs) to $E_8\times E_8$ heterotic string theory, and in general the hope has been
that the intricate mathematical structure of $E_8$ will `explain' or otherwise throw light on the structure of the Standard Model of Particle Physics (SMPP),
and/or point the way to extensions to the SMPP, and/or a unification with gravity, quantised or otherwise. These hopes remain largely unfulfilled,
and as a result the study of $E_8$ remains a fringe activity within mathematical and theoretical physics.

Perhaps the most famous example of an $E_8$ `Theory of Everything' (TOE), though by no means the first,
is Lisi's proposal \cite{Lisi}, which, however, attracted criticism \cite{DG} for two main perceived deficiencies: one was the lack of
a clear explanation for the chirality of the weak interaction, and the other was the lack of a convincing description of the three generations of elementary fermions.
Indeed, \cite{DG} went so far as to claim that these problems could not in principle be solved within $E_8$. However, their argument considers only a limited number of
possibilities, and does not fully take account of the richness of the structure of $E_8$. Their argument based on chirality switches between the complex and real forms of
$E_8$ in a manner which invalidates the argument \cite{chirality}, and their argument based on counting the number of Weyl spinors required assumes that the
symmetry of three generations requires a three-dimensional space, unfortunately ignoring the fact that an equilateral triangle in two dimensions has a perfectly good 
three-fold symmetry.

For these and other reasons, a number of people have continued to produce $E_8$ models \cite{Chester,octions}, but these models tend to
suffer from similar problems, typically finding that there is not quite enough room in $E_8$
for all the structure that seems to be required. In part, this problem arises because finite symmetries, such as the generation symmetry or the weak isospin symmetry,
do not exist in the Lie algebra formalism that is usually adopted. Thus a finite symmetry is always implemented with a copy of the Lie algebra $u(1)$, even if
the finite symmetry in the Lie group does not determine a copy of the Lie group $U(1)$. For example, both charge and weak hypercharge symmetries use copies of $U(1)$.

These copies of $U(1)$ introduce continuous variables that have no clear physical meaning, and are therefore superfluous.
  In this paper I shall show that by avoiding
the introduction of these superfluous variables, it is possible to embed all of the SMPP in $E_8$, in an essentially unique way. The model retains all the 
essential finite symmetries, including independent symmetries of order $3$ for (a) the three generations of electron and (b) the three quarks in a proton, 
or something similar, and 
including (c) a symmetry of order $2$ for weak hypercharge, and (d) a symmetry of order $4$ related to spin and (perhaps) photon polarisation.

\section{Overview of the model}
First I shall describe abstractly how the model is constructed, in the semi-split real form $E_{8(-24)}$ of $E_8$, and 
sketch a proof that, under certain reasonable assumptions,
this construction is unique.
Then I give a concrete description with explicit generators, and
demonstrate that the model reduces naturally to the SMPP.  
I then examine various aspects of the model to see what it predicts about physics beyond the standard model (BSM).

I start with a fundamental symmetry of order $3$, which will later be interpreted as a generation symmetry, for at least some, but not necessarily all, of the fundamental
fermions. The essential physical property of the $3$-generation symmetry for electrons is that
it is discrete, and does not arise from a continuous spectrum of possible electron masses. In the complex Lie group of type $E_8$, there are just two conjugacy classes
of subgroups $Z_3$ that do not extend canonically to a Lie group $U(1)$ with the same centralizer. The two centralizers are of types $A_2+E_6$ and $A_8$
respectively. In the compact real form, the latter group is precisely $SU(9)/Z_3$, but in other real forms the conjugacy class may split into different real forms of $A_8$.

In $E_{8(-24)}$ there is an element of order $3$ with centralizer $SU(7,2)/Z_3$. Within this group we need an element of order $2$ to be the central involution in the
Lorentz group, in order to define spin and spinors. Up to conjugacy, there is a unique possibility, and centralizing such an involution reduces the symmetry group to
\begin{align}
\label{genspin}
 (SU(5) \times SU(2,2)\times U(1))/Z_{20}
\end{align}
in which the $U(1)$ factor contains the $Z_3$ scalars
that will be interpreted as a generation symmetry for $10$ of the $15$ Weyl spinors in the Standard Model.
The $U(1)$ factor also provides a complex structure for both the gauge group $SU(5)$ and the conformal group $SU(2,2)\cong Spin(2,4)$, and their representations,
but otherwise plays no clear role in the SMPP. Omitting it leaves us with a simple direct product of three groups
\begin{align}
Z_3 \times SU(5) \times SU(2,2).
\end{align}

This group neatly combines the generation symmetry $Z_3$ with the gauge group of the Georgi--Glashow GUT \cite{GG} and the symmetry group
$SU(2,2)\cong Spin(4,2)$ of the Penrose twistors \cite{twistors,twistorlectures}. 
Hence restricting to the Standard Model from this point is quite straightforward.
We first split up $SU(5)$ by choosing either a symmetry of order $2$ (related to the weak interaction) or a symmetry of order $3$ (related to the strong interaction), to get the
Standard Model gauge group
\begin{align}
(SU(3) \times SU(2)\times U(1))/Z_6
\end{align}
Within $SU(5)$ both of these symmetries extend to a canonical copy of $U(1)$, but this is not true in $E_8$, where the symmetry of order $3$ has centralizer of
type $A_2+E_6$ and the symmetry of order $2$ has centralizer of type $A_1+E_7$.
It is also worth noting that by working with $U(5)$ rather than $SU(5)$, we can avoid the conventional quotient by $Z_6$, the necessity for which is questioned by Tong \cite{Tong}.

Finally we split $Spin(2,4)$ into
\begin{align}
Spin(1,1) \times Spin(1,3) 
\end{align}
in which Spin(1,1) provides a positive real variable which can be used as a Lorentz-invariant mass gauge, relating two real scalars to each other. 
The model is therefore described by the group
\begin{align}
SU(3)\times SU(2)\times U(1) \times Spin(1,1) \times Spin(1,3)\times U(1)
\end{align}
that augments the SMPP gauge group and `complexified' Lorentz group by adding (a) a generation symmetry group $U(1)=Spin(2)$ and (b) a mass gauge group $Spin(1,1)$. 
The interesting part of the model is therefore the way in which the generation symmetry couples to the mass gauge.
This coupling occurs via the embedding of $Spin(2)$ and $Spin(1,1)$ in a larger group, 
which will turn out to be $Spin(3,1)$, outside $SU(7,2)/Z_3$.

\section{Proof of uniqueness}
Lisi proposed \cite{Lisi} that the generation symmetry might be implemented as the discrete triality symmetry of a subgroup of
type $D_4+D_4$, rather than as a copy of $U(1)$, although this remained a vague aspiration rather than a concrete proposal.
But it turns out that this symmetry is not fundamentally discrete in the Lie group $E_8$, as its centraliser is $U(1)\otimes Spin(14)$.
This implies that there is no fundamental reason why there should be three generations, rather than four, or five, or any other number.
In order to classify the possibilities for a generation symmetry inside $E_8$, we must look at all elements of order $3$ in the group,
and determine which ones are fundamentally discrete, and which are not.

In fact, the elements of finite order in the (complex) Lie group $E_8$ are completely classified, and can be described inside a maximal torus.
Let us take coordinates for the root system so that the roots are all images under even coordinate permutations and even sign changes of
\begin{align}
&(1,1,0,0,0,0,0,0)\cr
&(1,1,1,1,1,1,1,1)/2.
\end{align}
There are four conjugacy classes of elements of order $3$, corresponding to the following elements of the torus
\begin{align}
&(1,0,0,0,0,0,0,0)\cr
&(1,1,0,0,0,0,0,0)\cr
&(1,1,1,0,0,0,0,0)\cr
&(1,1,1,1,1,0,0,0).
\end{align}

Let us call these elements of type 1, 2, 3 and 5 respectively.
The centralizers  are of type 
\begin{align}
&U(1)\otimes Spin(14),\cr 
&U(1)\otimes E_7,\cr 
&SU(3)\otimes E_6,\cr
&SU(9)/Z_3
\end{align}
 respectively.
In particular, the elements of types 1 and 2 extend to canonical copies of $U(1)$, which can have no physical meaning in the context of 
a discrete symmetry of three generations, or of three quarks in a proton. These elements cannot therefore be used for these discrete symmetries.
The elements of types 3 and 5, on the other hand, are fundamentally discrete, and cannot have any order other than $3$.
By embedding the torus in $Spin(12,4)$ we can choose each coordinate to correspond to a $2$-space in the $16$-space acted on by $SO(12,4)$,
so that calculations can be done in these groups rather than in $E_8$.
It is then easy to see that the centralizer of a commuting pair of one type $3$ and one type $5$ element is of type $SU(3)\otimes U(6)$.

In the compact real form of $E_8$, only compact real forms of these centralizers arise, but in non-compact real forms there can be a variety of different
real forms. The `semisplit' real form $E_{8(-24)}$ is generally considered to be the only possible real form to use for a fundamental physical theory.
In this case there is a subgroup $Spin(12,4)$ that contains the full torus, and we can easily see an element of type $5$
with centralizer containing
\begin{align}
SU(5) \times Spin(2,4) \cong SU(5) \times SU(2,2)
\end{align}
which implies that the full centralizer is $SU(7,2)/Z_3$.
Additionally centralizing a second element of type $3$, in such a way that the centralizer contains the Lorentz group $SL(2,\CC)$, results in the group
\begin{align}
(SU(3) \times SU(4,2)\times U(1))/Z_3,
\end{align}
in which the $SU(4,2)$ factor provides a context in which electro-weak unification can be described, independently of the strong force.

It remains to eliminate the possibility that the generation symmetry could be of type $3$. If it were, then it would centralize only a subgroup of type $A_2+A_2+A_2$
in $E_6$, so could only be Lorentz-invariant if two copies of $A_2$ combined to make the complex form $SL_3(\CC)$. 
This case does occur in the form $SU(3)\times SL_3(\RR)\times SL_3(\CC)$ in $E_{8(-24)}$, but does not allow for a Lorentz-invariant gauge group $SU(2)$ for the
weak interaction.
Thus this case is impossible, at least under most reasonable assumptions. 

\section{Explicit construction}
We use the notation of \cite{WDM,octions} for the Lie algebra of $E_{8(-24)}$ to describe an explicit copy of $SU(7,2)/Z_3$. The subgroup of the latter that lies in
$Spin(12,4)$ is a copy of the group given in (\ref{genspin}), and we begin by giving generators for its Lie algebra $u(5)+su(2,2)$.
For reasons that will become clear later, we choose $su(2,2)=so(2,4)$ to act on the labels $U,K,L,IL,JL,KL$ in $\OO'$, where $\OO'$ denotes the split octonion algebra, 
so that the algebra is generated by the $7$ rotations
\begin{align}
D_K,
D_{IL,L},D_{JL,L},D_{KL,L},D_{IL,JL},D_{JL,KL},D_{KL,IL}
\end{align}
and the $8$ boosts
\begin{align}
 D_L,D_{IL},D_{JL},D_{KL},D_{K,L},D_{K,IL},D_{K,JL},D_{K,KL}.
\end{align}
The scalar $u(1)$ we take to be generated by
\begin{align}
\label{mainu1}
D_{i,il}+D_{j,jl}+D_{k,kl}+D_{l}+D_{I,J} = D_{I,J}-2E_l
\end{align}
so that $su(5)$ is generated by the $24$ elements
\begin{align}
\label{su5gens}
&D_{l}- D_{i,il}, D_l-D_{j,jl}, D_l-D_{k,kl},D_l-D_{I,J}\cr
&D_{i}+D_{il,l}, D_{j,k}+D_{jl,kl}, D_{il}-D_{i,l}, D_{j,kl}+D_{k,jl},\cr
&D_{j}+D_{jl,l}, D_{k,i}+D_{kl,il}, D_{jl}-D_{j,l}, D_{k,il}+D_{i,kl},\cr
&D_{k}+D_{kl,l}, D_{i,j} + D_{il,jl}, D_{kl}-D_{k,l}, D_{i,jl} - D_{j,il},\cr
& X_{J}+X_{Il}, X_{Ji}+X_{Iil}, X_{Jj}+X_{Ijl}, X_{Jk}+X_{Ikl},\cr
& X_{Jl}-X_I, X_{Jil}-X_{Ii}, X_{Jjl}-X_{Ij}, X_{Jkl}-X_{Ik}.
\end{align}

Now to restrict from $SU(2,2)$ to the Lorentz group $SL(2,\CC)$ we pick a boost, such as $D_{K,L}$, to 
gauge the two scalars with the labels $K$ and $L$, which leaves $Spin(1,3)$ to act
on the labels $U,IL,JL,KL$, with generators for the Lie algebra: 
\begin{align}
D_{IL,JL}, D_{JL,KL},D_{KL,IL}, D_{IL}, D_{JL}, D_{KL}.
\end{align}
Similarly, to restrict from $su(5)$ to $u(3)+su(2)$ we pick a suitable rotation, such as $D_{i,il}+D_{j,jl}+D_{k,kl}$, to define the scalars of $u(3)$,
or equivalently we define the scalars of $u(2)$ with $D_{l}+D_{I,J}$. Either way we get generators for $su(2)$ as
 \begin{align}
 X_{I}+X_{Jl},X_{J}-X_{Il},  D_{I,J}-D_{l}
 \end{align}
 and generators for $su(3)$ as
 \begin{align}
 & D_{i,il}-D_{j,jl}, D_{j,jl}-D_{k,kl},\cr
& D_{j,k}+D_{jl,kl}, D_{j,kl}+D_{k,jl},\cr
& D_{k,i}+D_{kl,il},  D_{k,il}+D_{i,kl},\cr
& D_{i,j} + D_{il,jl},  D_{i,jl} + D_{j,il}.
 \end{align}

At this stage we have an embedding of the SMPP in $E_{8(-24)}$ that is almost identical to that described in \cite{chirality},
but with the labels $u=1$ and $K$ interchanged. We can therefore use much of the work done there, subject to minor changes in notation.
The embedding is also similar to that constructed in \cite{octions}, but differs in one important respect,
namely that the gauge group for the strong interaction is compact $SU(3)$, as in the standard model, rather than $SL(3,\RR)$, as in \cite{octions}.
At the same time, the Lorentz group switches from $Spin(3,1)$ to $Spin(1,3)$, which, as far as the SMPP is concerned, is a distinction without a difference,
but which, in the context of $E_{8(-24)}$, makes a real difference.

So far we have explicitly listed $40$ generators for $u(1) + su(5) + su(2,2)$, which covers the Standard Model and the well-known extensions to the
Georgi--Glashow GUT and Penrose twistors. There are a further $40$ generators for $su(7,2)$, which we first need to find explicitly. 
They lie in a complex tensor product of a $5$-dimensional representation of $SU(5)$ and a $4$-dimensional (twistor) representation of $SU(2,2)$,
so that according to the Georgi--Glashow classification they correspond
 to `left-handed' leptons ($16$ real elements) and `right-handed' down quarks ($24$ real elements). 
 The former are acted on trivially by the `right-handed' $U(1)$ generated by $D_{l}+D_{I,J}$, so include
 $Y_1-Y_{Kl}$ and  $Z_1+Z_{Kl}$.
The full list can then be obtained by applying the generators of $SU(2,2)$, which amounts to multiplying the $Y$ labels on the left, and the $Z$ labels on the right,
by $L$, $IL$, $JL$ and $KL$, repeatedly, giving
\begin{align}
&Y_1-Y_{Kl}, Y_I-Y_{Jl}, Y_J-Y_{Kl}, Y_K-Y_{Il},\cr
&Y_L+Y_{KLl}, Y_{IL}-Y_{JLl}, Y_{JL}-Y_{KLl}, Y_{KL}-Y_{ILl}\cr
&Z_1+Z_{Kl}, Z_I-Z_{Jl}, Z_J-Z_{Kl}, Z_K-Z_{Il},\cr
&Z_L+Z_{KLl}, Z_{IL}-Z_{JLl}, Z_{JL}-Z_{KLl}, Z_{KL}-Z_{ILl}.
\end{align}
In the Georgi--Glashow scheme, these represent the left-handed leptons, while the right-handed leptons have the opposite sign combinations.

We can then apply elements of $SU(5)$ to these to get the remaining $24$ elements of the adjoint representation of $SU(7,2)$. 
These elements represent the right-handed down quarks.
Each is the sum of four of the
remaining $96$ spinors ($Y$s and $Z$s), with particular sign combinations. I do not give a full list here, as it is cumbersome, 
and the particular signs combinations that occur are not very informative. 

\section{Generation symmetries}
However, a problem arises at this point, since the spinors (type $Y$ and $Z$) that lie inside $SU(7,2)/Z_3$ are not acted on by the proposed 
discrete generation symmetry. In the Georgi--Glashow model, these spinors represent the left-handed leptons and the right-handed down quarks (and their
anti-particles). To resolve this issue, we may try to use the fact that our model actually contains a canonical copy of $Z_3\times Z_3$, which contains four copies of $Z_3$.

To see this group $Z_3\times Z_3$ explicitly, we first split the generator (\ref{mainu1})
for the Lie algebra $u(1)$
into two pieces
\begin{align}
&D_{i,il}+D_{j,jl}+D_{k,kl},\cr
&D_{I,J}+D_{l},
\end{align}
and then exponentiate each one separately to give an element of order $3$. 
The first one is the central element of $SU(3)$, and has centralizer of the form $SU(3) \otimes E_6$.
It acts as a generation symmetry on all quarks, up and down, left-handed and right-handed, but no leptons.
The second one can be taken as a scalar in $U(2)=U(1)\otimes SU(2)$, and has full centralizer of the form $U(1) \otimes E_7$.
It acts as a generation symmetry on all right-handed particles, but no left-handed particles.
The sum of the two Lie algebra elements gives
the central $Z_3$ in $SU(7,2)/Z_3$, and the difference is another $Z_3$ in the same conjugacy class.
But in no case do we see an explicit generation symmetry on left-handed leptons.

Indeed, it is possible to split (\ref{mainu1}) up into individual terms, or any combinations, and all of them act as generation symmetries for certain types of
particles. Any single term is fixed-point-free on the $16$-dimensional complex spinor, so can be interpreted as a generation symmetry for all the particles,
left-handed and right-handed leptons and quarks. For example, the proposed generation symmetry in \cite{octions} is, after swapping $\OO$ with
$\OO'$ to convert $Spin(7,3)$ to compact $Spin(10)$,  equivalent to $D_{I,J}$.
However, this symmetry is not gauge-invariant, so distinguishes the electron generation symmetry from the neutrino generation symmetry,
and hence gives rise to the lepton-mixing (Pontecorvo--Maki--Nakagawa--Sakata \cite{Pontecorvo,MNS}
or PMNS) matrix. It also distinguishes the up quark generation symmetry from the down quark generation symmetry,
and hence gives rise to the Cabibbo--Kobayashi--Maskawa \cite{Cabibbo,KM} or CKM matrix. 
Moreover, it implies that the actual information content in the CKM and PMNS matrices is the same, 
as we shall demonstrate in Section~\ref{mixing}.

It may be worth also considering the group $Z_3\times U(1)$ as a gauge-invariant colour/generation
symmetry group, apparently including (some aspects of) weak hypercharge as well. Centralizing this group in $E_8$ is
$(SU(3)\times SU(4,2)\times U(1))/Z_3$, in which $SU(4,2)$ combines spin with weak isospin, and may therefore potentially provide
a mechanism for explaining the chirality of the weak interaction. Indeed, $SU(4,2)$ distinguishes clearly between the `left-handed' and `right-handed' copies
of $SU(2)$ in $SU(2,2)$, since the former combines with weak $SU(2)$ into $SU(4)$, and the latter does not. The centre of $SU(4)$ is then a group
$Z_4$ of order $4$, that behaves like a discrete version of the gauge group $U(1)_Y$ of weak hypercharge. Since weak hypercharge is indeed a discrete
concept, this copy of $Z_4$ is likely to be useful.
Another copy of $Z_4$ that is likely to be useful is the centre of $SU(2,2)$, that describes discrete properties of spin. Combining all these finite groups together gives a group
$Z_3\times Z_3 \times Z_4 \times Z_4$
that is fundamental to the model. 

\section{Properties of $SU(7,2)$}
In terms of the natural (defining) representation $\rep9$ of $SU(7,2)$, the adjoint representation of $E_{8(-24)}$ restricts as the adjoint
$\rep9 \otimes \rep9^*-\rep1$, with real dimension $80$, plus the exterior (or anti-symmetric) cube $\Lambda^3(\rep9)$, of complex dimension
$9.8.7/3.2.1=84$ or real dimension $168$. The complex structure of the latter is defined by the scalars of order $9$ in $SU(7,2)$,
which define the generation symmetry of (in particular) the right-handed electron. The mass hierarchy of the electrons then implies that complex
conjugation is not a symmetry of the model, and hence the model is unavoidably chiral.
The representation $\rep9$ itself does not appear in $E_8$, but is needed to describe the Standard Model in full, since it splits into
the colour representation $\rep3$ of $SU(3)$, plus the weak doublet representation $\rep2$ of $U(2)$, plus the twistor or Dirac spinor representation $\rep4$ of $SU(2,2)$.

In order to describe explicit elements of $SU(7,2)$ we adopt the usual physicists' convention of representing elements of the Lie algebra by
Hermitian matrices, with an implied factor of $2\pi i$ inserted before exponentiation into the Lie group. In particular, the scalars of order $9$ are 
represented by the diagonal matrix $(1,1,1,1,1,1,1,1,1)/9$, and its multiples by $\pm1, \pm2, \pm4$. The requirement for such an element to lie in $SU(7,2)$ is simply
that the trace is an integer. Continuous symmetries must have trace $0$, so that all scalar multiples exponentiate to determinant $1$, but discrete symmetries
do not have this restriction. The central element $(1,1,1,1,1,1,1,1,1)/3$ with trace $3$
acts trivially in $E_8$, so that for some purposes we can work with the trace modulo $3$.

The central element of $SU(3)$ is represented by $(1,1,1,0,0,0,0,0,0)/3$, 
so that the four copies of $Z_3$ in $Z_3\times Z_3$ are represented by
\begin{align}
&(1,1,1,0,0,0,0,0,0)/3\cr
&(1,1,1,1,1,1,1,1,1)/9\cr
&(4,4,4,1,1,1,1,1,1)/9\cr
&(7,7,7,1,1,1,1,1,1)/9
\end{align}
Of these, the first three are fundamentally discrete, since the trace is non-zero modulo $3$, while the last extends to $U(1)$,
since the trace is $3$, that is equivalent to zero on re-writing $\exp(2\pi i(7/9))$ as $\exp(2\pi i(-2/9))$. 

The scalar $4/9$ can be extended to $u(1)$ by re-writing it as
\begin{align}
(4,4,4,4,4,-5,-5,-5,-5)/9
\end{align}
which has the effect of splitting $SU(7,2)$ into $SU(5)\times SU(2,2)$, with a copy of $U(1)$ to provide the scalar multiplications on the fundamental
representations of both factors. The latter are obtained by multiplying by $9/5$ and $9/4$ respectively,
to obtain
\begin{align}
(4,4,4,4,4,-5,-5,-5,-5)/5 & \equiv -(1,1,1,1,1,0,0,0,0)/5,\cr
(4,4,4,4,4,-5,-5,-5,-5)/4 & \equiv -(0,0,0,0,0,1,1,1,1)/4
\end{align}
respectively.
In other words, this procedure gives us another way of looking at the symmetry-breaking down to Georgi--Glashow plus twistors.
Moreover, since we now have two copies of $U(1)$, we break the symmetry further to the Standard Model gauge group, plus twistors.
In effect what has happened is that we have taken three finite symmetries, two of order $3$ and one of order $4$, that are fundamental to 
particle physics, and extended them to $U(1)$. But since there are only two copies of $U(1)$, and three finite symmetries,
any particular choice of generators for the former will lead to mixing of these generators for the latter.
\section{Mixing angles}
\label{mixing}
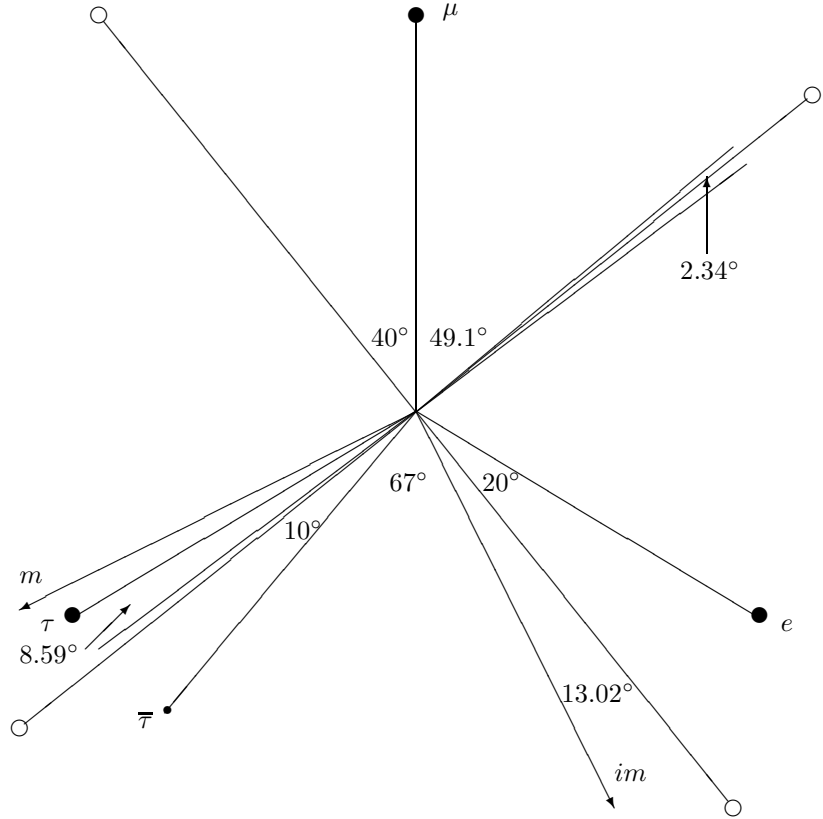
\begin{figure}
\caption{\label{mixfig}Schematic diagram of mixing angles}
\begin{picture}(300,320)
\put(150,150){\line(5,-3){130}}
\put(150,150){\line(0,1){150}}
\put(150,150){\line(-5,-3){130}}
\put(150,150){\line(4,-5){118}}
\put(150,150){\line(-4,5){118}}
\put(150,150){\line(5,4){148}}
\put(150,150){\line(-5,-4){148}}
\put(150,150){\line(6,5){120}}
\put(150,150){\line(4,3){125}}
\put(150,150){\line(-4,-3){120}}
\put(20,73){\circle*{6}}
\put(150,300){\circle*{6}}
\put(280,73){\circle*{6}}
\put(270,0){\circle{6}}
\put(30,300){\circle{6}}
\put(300,270){\circle{6}}
\put(0,30){\circle{6}}
\put(160,300){$\mu$}
\put(288,67){$e$}
\put(8,67){$\tau$}
\put(150,150){\vector(-2,-1){150}}
\put(150,150){\vector(1,-2){75}}
\put(0,85){$m$}
\put(133,175){$40^\circ$}
\put(155,175){$49.1^\circ$}
\put(205,40){$13.02^\circ$}
\put(225,10){$im$}
\put(0,55){$8.59^\circ$}
\put(25,60){\vector(1,1){17}}
\put(175,120){$20^\circ$}
\put(250,200){$2.34^\circ$}
\put(260,210){\vector(0,1){29}}
\put(150,150){\line(-5,-6){95}}
\put(140,120){$67^\circ$}
\put(100,102){$10^\circ$}
\put(45,30){$\overline\tau$}
\put(56,37){\circle*{3}}
\end{picture}
\end{figure}

The finite scalars of order $3$ in $SU(3)$, of order $4$ in $SU(2,2)$, and of order $9$ in $SU(7,2)$, as well as those of order $2$ in weak $SU(2)$,
 then combine to give a
plethora of mixing angles in the SMPP, as we proceed to demonstrate.  The full picture is illustrated in Figure~\ref{mixfig}, which I shall explain step by step.
First of all, combining the scalar $-5/9$ on the twistors with the natural scalar $1/4$, or with the natural scalar $1/2$ for spinors, divides the circle into units of $10^\circ$ or $20^\circ$,
which have fundamental physical meanings in terms of relating generation symmetries to the complex scalars of the Standard Model. In particular, the scalar
$-5/9$ corresponds to an angle of $200^\circ$, and $1/2-5/9=-1/18$ is an angle of $20^\circ$. Similarly $1/4-2/9$ is $10^\circ$ and $1/4-1/9$ is $50^\circ$.

These angles, or others like them, should therefore be expected to appear in or related to the CKM \cite{Cabibbo,KM}
and/or PMNS matrices \cite{Pontecorvo,MNS}. They clearly do not appear in the CKM matrix, but the
four angles in the PMNS matrix are around $33^\circ$, $8.5^\circ$, $49^\circ\pm1^\circ$ and $197^\circ\pm 25^\circ$, which suggests 
that there might be a close relationship between
the last three and $10^\circ$, $50^\circ$ and (possibly) $200^\circ$ respectively. The discrepancies between the exact values and the
experimental values are, however, in at least one case, too large to be explained by experimental
uncertainty, so we must conclude that the PMNS angles are also not exact multiples of $10^\circ$.

In the remaining case, the corresponding entry in the CKM matrix is the Cabibbo angle of $13^\circ$, so that the difference between the two is now $20^\circ$.
This suggests that the discrepancies in the other two cases, totalling around $1.5+.9=2.4^\circ$, may be related to the other angles in the CKM matrix, of
around $2.4^\circ$ and $0.2^\circ$, distributed in some way to be explained. Now the explanations in \cite{finite} for the two angles of $33^\circ$ and $2.38^\circ$
involve four-dimensional quaternion geometry/algebra, which is hard to visualise, but a good idea can still be obtained by projecting onto a complex plane,
with $U(1)$ acting on it, including $Z_3$ for a generation symmetry, $Z_4$ for the internal workings of the weak force,
and $Z_9$ for the mixing of two coordinate systems. 

There is an overall twist of $200^\circ$ between $Z_3$ and $Z_4$, or $20^\circ$ plus a sign change.
The directions defined by $Z_3$ then differ from the four directions defined by $Z_4$ by $20^\circ$ and $200^\circ$ ($\pm1$ in $Z_4$ matched to $+1$ in $Z_3$) and
$10^\circ$ and $50^\circ$ ($\pm i$ in $Z_4$ matched to $\exp(\pm2\pi i/3)$ in $Z_3$). The measured Cabibbo angle of $13.02^\circ$ is added to the $20^\circ$ twist to
produce the angle of $33.02^\circ$ obtained in \cite{finite} from the mass ratios of the three generations of electrons. 
The next largest angle in the CKM matrix is measured
at around $2.38^\circ$, and was calculated as $2.337325^\circ$ in \cite{finite} by adding in the proton mass to the system of equations.
If we assume that this angle relates to the other $Z_4$ axis, that is $\pm i$, then it projects onto the other two entries in the PMNS matrix in the ratio
$\cos10^\circ$ to $\cos50^\circ$, that is around $1.532$, giving a splitting into $1.414^\circ+.923^\circ$. This procedure adjusts the exact values to
\begin{align}
10^\circ - 1.414^\circ & = 8.586^\circ,\cr
50^\circ - .923^\circ  & = 49.077^\circ
\end{align}
in complete agreement with the measured values
\begin{align}
{8.54^\circ}^{+.11^\circ}_{-.12^\circ},\cr
{49.1^\circ}^{+1.0^\circ}_{-1.3^\circ}.
\end{align}

See Figure~\ref{mixfig} for an illustration of these angles in relation to the square (open circles) and the equilateral triangle (filled circles),
to show how the interplay between $Z_3$ and $Z_4$ works. Also shown is the mass axis
for electrons (marked $m$), and its perpendicular (marked $im$, since the picture represents a complex plane). 
Note that there are three lines whose position is measured by experiment, that is the imaginary mass axis $im$, and the two lines very close together near the diagonal
of the square. The PMNS angles are the angles between these three lines and the altitudes of the triangle. The Cabibbo angle is the angle between $im$ and
a diagonal of the square, and the $(2,3)$ mixing angle of the CKM matrix is the angle between the other two experimental lines. Notice also that the diagram clearly shows the
lepton generation symmetry, but does not clearly show the quark generation symmetry. The latter uses a different copy of $Z_3$, which acts perpendicularly to the page,
so cannot be seen in this projection.

This leaves just one mixing angle for which no potential explanation has been offered here or in \cite{finite}, namely the down/bottom quark mixing angle of $.201\pm.011$.
Such a small angle can easily disappear in the experimental noise, so that no clues can be expected from the experimental values of the other mixing angles.
Only the proposed calculated values \cite{finite} are sufficiently precise to
provide any meaningful clues. Such clues might lead to a conjecture for this last mixing angle, although it will ultimately be necessary to provide a mathematical justification.

We may perhaps be struck by the fact that the sum of the
electron/muon neutrino mixing angle and the CP-violating phase of the CKM matrix evaluates to $99.75^\circ$, whose difference from $100^\circ$
is of the right order of magnitude. As in the other case above, we should expect the $10^\circ$ rotation from $90^\circ$ to $100^\circ$ to have an effect on
the measured value of the angle, and it may be reasonable to conjecture that $\cos100^\circ\approx -.17365$ enters into the calculation somehow.
Reducing the angle of $.25^\circ$ by this proportion results in an angle of $.21^\circ$, in line with experiment. However, there is no theoretical justification for this
calculation, which remains purely conjectural. 

Nevertheless, some of this discussion amounts to a genuine prediction of 
more accurate values of many of the mixing angles.
These predictions arise partly from the realisation that the CKM and PMNS matrices are not independent.
What this discussion also 
explains is 
why the neutrino-mixing angles are not small, like the quark-mixing angles. Finally, it possibly indicates that the CP-violating phase in the PMNS matrix may not be
exactly $180^\circ$, as some people have suggested, but instead exactly $200^\circ$.

\section{Electro-weak mixing}
The splitting of $SU(2)\times SU(2,2)$ into three factors
\begin{align}
SU(2)_W \times SU(2)_L \times SU(2)_R
\end{align}
begs the question of how the association of the weak $SU(2)_W$ with the left-handed $SU(2)_L$ comes about. Any two of the 
copies of $SU(2)$ can be combined,
to give 
\begin{align}
SU(2)_W \times SU(2,2)_{LR},\cr
SU(2)_L \times SU(2,2)_{WR},\cr
SU(2)_R \times SU(4)_{WL}.
\end{align}
Each of these groups has centre $Z_2\times Z_4$, giving three different copies of $Z_2$ and $Z_4$, together with various `mixed' copies. 
The question then is how to interpret all these finite symmetries. 

In \cite{finite}, a version of the Weinberg angle that is independent of energy 
is identified as a mixing angle between a copy of $Z_2$ representing charge conjugation, and a $Z_4$ which acts by swapping 
weak hypercharge with the third component of weak isospin. 
If these are Lorentz-invariant, they must be the central elements of $SU(2)_W$ and $SU(2,2)_{LR}$ respectively.
 However, the actual Weinberg angle measured in experiments `runs' with the energy scale, so that the latter must be replaced by $SU(4)_{LW}$.
 If this is correct, the Weinberg angle describes a mixing between $SU(2)_W$ and $SU(2)_L$,
 as we would expect from the chirality of the weak interaction. Then we obtain a Weinberg angle that is invariant under $SU(2)_R$, but not
necessarily under the Lorentz group $SL(2,\CC)$. Hence the model allows (or perhaps requires) the Weinberg angle to `run' with the energy scale.

The same may also be true for (some of) the other mixing angles discussed above, since they also relate to a copy of $Z_4$, and it is not
immediately obvious which copy of $Z_4$ is most appropriate. Only the $LR$ copy of $Z_4$ is Lorentz-invariant, so that if we use either the
$WL$ copy or the $WR$ copy, then the angles should again `run' with the energy scale. This fact again illustrates the advantages of using
the theory of twistors to split $SU(2)_L$ from $SU(2)_R$, rather than relying solely on the representation theory of the Lorentz group, since the
latter approach naturally produces mixing angles that do not run with energy.

It may also be possible to identify the parity ($P$) and/or time reversal ($T$) symmetries here. I have already suggested the central involution $-1_W$
in $SU(2)_W$ for charge conjugation ($C$), while $-1_L$ and $-1_R$ are rotation-invariant but not Lorentz-invariant, just like the $P$ and $T$ symmetries.
In fact, they both act on Minkowski space as $P$, while the scalar $i$ acts as $PT$. 
Hence the usual language of $C$, $P$ and $T$ does not capture the full structure here, since it ignores the difference between $-1_L$ and $-1_R$,
which is in reality the most important distinction we need to make. It is a fundamental
difference between compact $SU(4)_{WL}$ and non-compact $SU(2,2)_{WR}$, that creates the fundamental chirality of the weak interaction.

In particular, the usual assumption of $CPT$ invariance becomes instead the assumption of $-1_{WLR}$ invariance. This assumption implies that
the model is covariant under the group $SU(4,2)$ that centralizes $-1_{WLR}$. Hence it is within the group $SU(4,2)$ that we must describe the phenomenon
that is usually called $CP$-violation.

\section{Twistors}
Both Penrose twistors and Dirac spinors are elements of complex $4$-space, and there is a sense in which these two spaces can be
identified, so that Penrose twistors and Dirac spinors become just different ways of looking at the same thing. The Dirac algebra of all
$4\times 4$ complex matrices not only has its Standard Model interpretation as the complex Clifford algebra $\CC\ell(3,1)$ of Minkowski space,
but also as the even part of either or both real Clifford algebras $C\ell(4,2)$ and $C\ell(2,4)$. The Dirac $\gamma$ 
matrices then acquire an interpretation
as generators for the Lie algebra $so(2,4)=su(2,2)$.

The standard conventions for Dirac matrices, both the Bjorken--Drell \cite{BjorkenDrell} and chiral conventions, obscure these facts, so I use a different convention,
as follows.  First we take the Pauli matrices
\begin{align}
\sigma_1=\begin{pmatrix}0&1\cr 1&0\end{pmatrix}, \quad
\sigma_2=\begin{pmatrix}0&-i\cr i&0\end{pmatrix}, \quad
\sigma_3=\begin{pmatrix}1&0\cr 0&-1\end{pmatrix}, \quad
\end{align} 
and define the Dirac matrices in terms of them
\begin{align}
&\gamma_1=\begin{pmatrix}i\sigma_1&0\cr 0&-i\sigma_1\end{pmatrix}, \quad
\gamma_2=\begin{pmatrix}i\sigma_2&0\cr 0&-i\sigma_2\end{pmatrix}, \quad
\gamma_3=\begin{pmatrix}i\sigma_3&0\cr 0&-i\sigma_3\end{pmatrix}, \cr
&
\gamma_0=\begin{pmatrix}0&i\cr -i&0\end{pmatrix}, \quad\gamma_5=\begin{pmatrix}0&1\cr 1&0\end{pmatrix}.
\end{align}
One can also swap $\gamma_0$ and $\gamma_5$ here, in the same way that they are swapped between the Bjorken--Drell and
chiral conventions in the Standard Model.

The compact part of this algebra is $so(2)+so(4)$, in which $so(2)$ is generated by $\gamma_0\gamma_5$, and $so(4)$ by
$\gamma_1,\gamma_2,\gamma_3$ and
\begin{align}
\gamma_1\gamma_2=\begin{pmatrix}i\sigma_3 & 0\cr 0 & i\sigma_3\end{pmatrix},\quad
\gamma_2\gamma_3=\begin{pmatrix}i\sigma_1 & 0\cr 0 & i\sigma_1\end{pmatrix},\quad
\gamma_3\gamma_1=\begin{pmatrix}i\sigma_2 & 0\cr 0 & i\sigma_2\end{pmatrix}.
\end{align}
This shows that the algebra $so(4)$ splits into a left-handed part generated by 
\begin{align}
\gamma_1+\gamma_2\gamma_3, \gamma_2+\gamma_3\gamma_1, \gamma_3+\gamma_1\gamma_2,
\end{align}
 and a right-handed part generated by 
 \begin{align}
 \gamma_1-\gamma_2\gamma_3, \gamma_2-\gamma_3\gamma_1, \gamma_3-\gamma_1\gamma_2.
 \end{align}

 The $8$ boosts include $\gamma_0$ and $\gamma_5$, and their multiples with $\gamma_1,\gamma_2,\gamma_3$, so that the Lorentz group
 adds the generators
 \begin{align}
 \gamma_0\gamma_1 = \begin{pmatrix}0&\sigma_1\cr \sigma_1 & 0\end{pmatrix},\quad
 \gamma_0\gamma_2 = \begin{pmatrix}0&\sigma_2\cr \sigma_2 & 0\end{pmatrix},\quad
 \gamma_0\gamma_3 = \begin{pmatrix}0&\sigma_3\cr \sigma_3 & 0\end{pmatrix},
 \end{align}
 to the three rotations $\gamma_1\gamma_2$, $\gamma_2\gamma_3$ and $\gamma_3\gamma_1$,
 and the remaining three boosts are
  \begin{align}
 \gamma_5\gamma_1 = \begin{pmatrix}0&-i\sigma_1\cr i\sigma_1 & 0\end{pmatrix},\quad
 \gamma_5\gamma_2 = \begin{pmatrix}0&-i\sigma_2\cr i\sigma_2 & 0\end{pmatrix},\quad
 \gamma_5\gamma_3 = \begin{pmatrix}0&-i\sigma_3\cr i\sigma_3 & 0\end{pmatrix},
 \end{align}
 The Lorentz group here commutes with $\gamma_5$, which generates $so(1,1)$, that is a copy of the real numbers.
  The other copy of $so(3,1)$, involving $\gamma_5$, commutes with $\gamma_0$, which is related to energy in the Dirac equation.
 
 Translating into the language of $E_{8(-24)}$, with $SO(2,4)$ acting on the labels $U,K,L,IL,JL,KL$, we can take 
 \begin{align}
 \gamma_0 = D_L, \quad \gamma_1 = D_{L,IL},\quad \gamma_2=D_{L,JL}, \quad \gamma_3=D_{L,KL},\quad \gamma_5= D_{K,L}
 \end{align}
 This allows us to write down a Dirac equation, as soon as we have identified a suitable complex scalar $i$ for the mass term. The previous discussion
 encompasses various copies of $U(1)$ which mix together in various ways, but for the purpose of constructing a Dirac equation we need an element that
 `naturally' has order $4$. The obvious choice is the centre of $SU(2,2)$, which is (up to an arbitrary sign) 
 the product of the five gamma matrices, and therefore agrees with
 the standard choice of scalar in the Dirac algebra. 
 This can be written as the product of $\gamma_0\gamma_5$, $\gamma_1\gamma_2$ and $\gamma_3$ and hence as the product of
 $D_K$, $D_{IL,JL}$ and $D_{KL,L}$. Translating to additive notation for the Lie algebra allows us to use $D_K+D_{IL,JL}+D_{KL,L}$
 to represent this scalar.

The main thing that twistors do for us is to provide a method of converting from Minkowski spacetime (as required by relativity theory)
to Euclidean spacetime (as required by quantum field theory) without the artificial introduction of complex scalars, and without the resulting
complexities of Wick rotation. The conversion is simply achieved by replacing the generators 
$\gamma_0\gamma_1, \gamma_0\gamma_2, \gamma_0\gamma_3$
for the Lorentz group by the generators
\begin{align}
\gamma_1,\gamma_2,\gamma_3
\end{align}
for the Euclidean isometry group, rather than by multiplying $\gamma_0$ by $i$. As a result, chirality becomes a simple matter of splitting
up a real Lie group
\begin{align}
SO(4) = SU(2)_L \otimes  SU(2)_R
\end{align}
rather than splitting the corresponding complex Lie group into two copies of the Lorentz group, whose relationship to the real Lorentz group is at best obscure.
It then becomes possible to implement the chirality of the weak interaction by a suitable merger
between the gauge group $SU(2)_W$ and the left-handed spin group $SU(2)_L$.
In order to see how this might work, it is first necessary to look at the allocation of spinors to particles.

\section{Particles}
Next we consider how the adjoint representation breaks up under the Georgi--Glashow--Penrose algebra $su(5) + su(2,2) + u(1)$. The $u(1)$ component provides
complex structures for all the representations except the adjoint $\rep{24}+\rep{15}+\rep1$. 
The decomposition of the complex adjoint representation of $su(7,2)$ is given by
\begin{align}
(\rep5+\rep4)\otimes(\rep5^*+\rep4^*) -\rep1 = \rep{24}+\rep{15}+\rep1 + \rep5\otimes\rep4^* + \rep5^*\otimes\rep4
\end{align}
from which we have to choose one of the two complex structures, say $\rep5^*\otimes \rep4$.
The complex structure here is defined by exponentiating $(4,4,4,4,4,-5,-5,-5,-5)\theta/9$.
The remainder of the representation is
\begin{align}
\Lambda^3(\rep5+\rep4) & = \Lambda^3(\rep5) + \Lambda^2(\rep5)\otimes\rep4 + \rep5\otimes \Lambda^2(\rep4) + \Lambda^3(\rep4)\cr
& = \rep{10}^* + \rep{10}\otimes\rep4 + \rep5\otimes\rep6 + \rep4^*
\end{align}
again with the same copy of $U(1)$ to define the complex structure.
It should be noted, however, that this copy of $U(1)$ also contains the central $Z_5$ in $SU(5)$, which does not appear in the Standard Model at all,
and does not have any obvious physical meaning, since five-fold symmetries do not play any role in the SMPP. It is therefore not clear that this is the `correct'
copy of $U(1)$ to use for the scalars. It may be that $(-2,-2,-2,1,1,1,1,1,1)\theta/9$ does a better job.

The fermionic part of the representation is then
\begin{align}
\rep5^*\otimes\rep4 + \rep{10}\otimes\rep4 + \rep1\otimes\rep4^*,
\end{align}
which looks like the Georgi--Glashow scheme, including a right-handed neutrino, but with Penrose twistors instead of Weyl spinors. This extension to twistors is
required in order to have a non-trivial generation symmetry, but at the same time does not play well with the conventional division into left-handed and right-handed
particles, since each twistor consists of one left-handed and one right-handed Weyl spinor. 
The distinction between left-handed and right-handed twistors, $\rep4$ and $\rep4^*$, does not seem to help much either, even if we change our minds about the
complex structure on the first component, and write $\rep5\otimes\rep4^*$ instead.

Part of the problem may be that we have two different definitions of the scalar $U(1)$, which agree on the twistor representation, but not on the gauge group representations.
We can combine these scalars at arbitrary angles, three of which are of particular interest, since they act trivially on one of the representations:
\begin{align}
&(0,0,0,2,2,-1,-1,-1,-1),\cr
&(4,4,4,0,0,-3,-3,-3,-3),\cr
&(2,2,2,-3,-3,0,0,0,0).
\end{align}
The first of these links the scalar $i$ on the spinors and twistors to the scalar $-1$ on weak isospin and/or weak hypercharge, and is therefore of particular interest
in the context of electro-weak unification.  This linking of $Z_2$ to $Z_4$ was used in \cite{finite} to calculate an accurate value for the Weinberg angle.
The other two link $Z_3$ to $Z_2$ and $Z_4$, and are therefore important for describing the mixing angles between the three generations,
as described above.

The other problem is that we have two conflicting definitions of left-handed and right-handed, one of which splits the twistor into two spinors
\begin{align}
\rep4 & \rightarrow \rep2_L+\rep2_R
\end{align}
and the other of which splits the forces into weak (chiral) and strong (non-chiral)
\begin{align}
\rep5 & \rightarrow \rep2_W + \rep3,\cr
\rep{10} & \rightarrow \rep1 + \rep3^* + \rep2_W\otimes\rep3 
\end{align}
The formalism of the SMPP associates the two representations $\rep2_L$ and $\rep2_W$ with each other, in order to model the chirality of the weak interaction.
The model under consideration here does not have such an association, so either it has to be introduced, or we have to explain chirality in some other way.

The essential point is to introduce a group of some kind that separates $2_L$ from $2_R$. We can use the copy of $U(1)$ defined by
\begin{align}
(0,0,0,1,1,1,1,-2,-2)
\end{align}
to do this, or we can use the finite symmetry of order $4$ defined by
\begin{align}
(0,0,0,1,1,1,1,0,0)/4,
\end{align}
in which case we have broken the symmetry of $SU(2,2)$ down to the compact group 
$SU(2)_L \times SU(2)_R$. If we want instead to use the standard splitting into left-handed
and right-handed Weyl spinors, we have to use a copy of $GL(1,\RR)$ instead. Either way, we have a concept of `left-handed' spinor that is associated to the weak force,
and a concept of `right-handed' spinor that is not.

A possible clue to how to proceed comes from the fact that the Georgi--Glashow model mixes quarks with leptons, and predicts proton decay, which has never been observed.
Perhaps, therefore, a slightly different interpretation of the representations (as particles, or otherwise) is required. 
The diagram above illustrates the general method for combining a $Z_3$ generation symmetry with a $Z_4$ symmetry, that relates
left-handed and right-handed Weyl spinors via the scalar multiplication on twistors. However, this diagram is really a projection from $4$-dimensional space (quaternions)
onto $2$-dimensional space (complex numbers), as shown in \cite{finite}, so that the full picture of elementary particles is likely to be rather more complicated than the
diagram suggests.

\section{The electron}
\label{electron}
The standard extension of the $SU(5)$ GUT to $Spin(10)$ embeds naturally in $E_{8(-24)}$, 
and throws some light on the splitting of the electron into left-handed and right-handed parts.
The complex representations $\rep1$, $\rep5$ and $\rep{10}$ of $SU(5)$ combine into a complex
$16$-dimensional representation of $Spin(10)$. 
This larger group contains elements that unite the left-handed and right-handed parts of the particles,
effectively converting two copies of the complex numbers into a single quaternion.
This quaternion structure was used in an essential way in \cite{finite} to calculate the
Weinberg (electro-weak) mixing angle from the masses of the three electrons and the proton.

In order to embed this quaternion structure in $E_{8(-24)}$, we have to extend $Spin(10)$ to $Spin(11)$, and in order to
add masses to this quaternion structure, we have to extend further to $Spin(11,1)$. This is the full centralizer of the Lorentz group in $E_{8(-24)}$,
and therefore it must contain everything that is Lorentz-invariant, including all the fundamental masses.
Since $Spin(11,1)$ contains just $11$ independent boosts, this is the number of independent masses we can accommodate, splitting $6+5$ between quarks
and leptons. However, one of the boosts acts as a mass gauge group, so reduces the number of independent masses to $10$. The evidence from \cite{finite}
is that we can take these $10$ masses to be six quarks, three electrons and the proton, with the neutron acting as the mass gauge.

Thus the fundamental spin representation that describes the particles becomes a set of $8$ quaternions, representing $8$ particles,
namely the electron, neutrino and three colours of up and down quarks. The generation symmetry as originally described acts by
complex multiplication by $Z_3$ on $\rep1$ and $\rep{10}$, and acts trivially on $\rep5$. This element, incidentally,
lies inside $Spin(10)$, so that in a certain sense the $Spin(10)$ GUT already contains a generation symmetry. There is perhaps an
unexpected feature, however, that the generation symmetry acts on right-handed leptons, left-handed down quarks, and all up quarks,
but not on left-handed leptons or right-handed down quarks. 
This is somewhat surprising, especially in the case of the neutrinos, and
requires investigation. The crucial point is to understand how the splitting of the quaternions into left-handed and right-handed complex parts relates to
the splitting of the Penrose twistor (Dirac spinor) into left-handed and right-handed Weyl spinors. 

Now the quaternion structure of the electron (three generations) was studied in some detail in \cite{finite}, where coordinates were chosen so that
the three generations were
\begin{align}
e & = -1+i+j\cr
\mu & = -1+j+k\cr
\tau & = -1+k+i
\end{align}
and rotation of $i$ to $j$ to $k$ to $i$ effects the generation symmetry.
It was then found  \cite{finite,remarks} that by including the proton and neutron as well
\begin{align}
p & = 1+i+j+k\cr
n& = i+j+k
\end{align}
there was a linear relation
\begin{align}
e+\mu+\tau + 3p = 5n
\end{align}
which 
is an exact equality of both charge and mass (as measured experimentally).

There is therefore a mass axis within the quaternions that lies at certain specific angles to certain particles or linear combinations of particles.
In \cite{finite} it was shown that three of these angles can plausibly be identified as the Weinberg (electro-weak mixing) angle, the
mixing angle $\theta_{sb}$ between strange and bottom quarks (fixing the up and down quarks that are in the proton and neutron), and the
CP-violating phase of the CKM matrix.

The splitting into left-handed and right-handed parts is determined by the generation symmetry, which acts on the right-handed parts but not the
left-handed parts. Hence we have
\begin{align}
e_L=\mu_L=\tau_L & = -1 +2(i+j+k)/3\cr
e_R & = (i+j-2k)/3\cr
\mu_R & = (j+k-2i)/3\cr
\tau_R & = (k+i-2j)/3
\end{align}
Moreover, the corresponding complex subalgebra of the quaternions is generated by $i+j+k$. This enables us to put the right-handed electrons
on the vertices of an equilateral triangle, project the mass axis onto this plane, and calculate another mixing angle from it, as is done in \cite{finite}.
This angle is plausibly identified with the mixing angle $\theta_{e\mu}$ between electron and muon neutrinos in the PMNS matrix,
since the quaternionic structure of the neutrinos (plus the neutron) is essentially the same as that of the electrons (plus the proton).

As we have seen above,
the same basic structure can be transferred also to the down quark representations (but not to the up quarks), 
with an additional rotation of the complex numbers through an angle of
$5/9-1/2$ of a rotation, to give the Cabibbo angle $\theta_{ds}\approx 13.02^\circ$ correct to $1/100$ of a degree.

Now we need to look at the tensor products of these quaternions with the spinors, in order to embed the particles in spacetime. At this point
we find a big difference between the neutrinos and the electrons, essentially because the electrons lie in $\rep5\otimes \rep4^*+\rep{10}\otimes \rep4$,
so that there is a clear separation into $\rep4+\rep4^*$, while the neutrinos lie in $\rep5\otimes \rep4^* +\rep1\otimes \rep4^*$ so do not make the same distinction.
That is, the $Spin(10)$ electrons use $\rep4+\rep4^*$ to split right-handed from left-handed, while the $Spin(10)$ neutrinos are all (left-handed) $\rep4^*$.

\section{Neutrinos}
In other words, the interpretation of $\rep1$ as `right-handed' neutrinos in the $Spin(10)$ model is not consistent with the representation theory, since this representation has a
`left-handed' relationship to spin. The splitting of $\rep4^*$ into $\rep2^*_L+\rep2^*_R$ is then a splitting into (left-handed) neutrinos and (right-handed) antineutrinos.
The same splitting defines the electrons and the anti-electrons, which are similarly defined as the `same' particle, but with opposite embeddings in spacetime.

In this discussion we see again the fundamental distinction between the handedness of twistors, and the handedness of Weyl spinors, that was discussed at length
in \cite{chirality}. This distinction does not exist in the Standard Model, but is a central feature of $E_8$ models. The fact that $E_8$ models do not reproduce the
details of the Standard Model at this point is used by some \cite{DG} as an argument that $E_8$ models cannot correctly model the chirality of the weak interaction.
I argue instead that chirality is more subtle than the Standard Model appreciates \cite{chirality}.

This begs the question of what is the difference between the neutrinos that lie in $\rep5$ and the neutrinos that lie in $\rep1$? 
Both of them are `left-handed' in the twistor sense, and the handedness of the Weyl spinors merely distinguishes between neutrinos and anti-neutrinos.
The neutrinos in $\rep5$ participate in the
weak interaction, but do not have a generation label, while the neutrinos in $\rep1$ have a generation label, but do not participate in the weak interaction. 

If they do not participate in the
weak interaction, then what do they participate in? The only possibility is that they participate in gravity. But we must not forget that they are not particles, they are only
\emph{projections} of particles onto a `right-handed' mathematical space. The actual particles participate in both the weak interaction \emph{and} gravity,
in such a way that the `right-handed' gravitational interaction affects the generation, but the `left-handed' weak interaction does not.
In other words, gravity itself is responsible for neutrino oscillations.

This idea should not really come as a surprise, since we have already deduced the neutrino-mixing angle $\theta_{e\mu}$ from the masses of the
three generations of electron. The other neutrino-mixing angles appear to come from adding in the proton and neutron masses, together with some
discrete geometry of quaternions, and suitable projections onto complex numbers. Hence we have claimed that these three independent mass ratios
are sufficient to explain all the observed neutrino oscillations quantitatively. It is therefore not a huge leap to assume that the physical mechanism by which the masses
determine the mixing angles is (quantum) gravity. %There is certainly no other known `force' that could do it.

We can separate the neutrinos from the other particles by restricting from $SU(5)$ to $SU(4)$, so that $Spin(2,4)$ extends to $Spin(4,4)$. The latter contains a subgroup
$GL_4(\RR)$ that represents spacetime coordinate changes, while preserving the duality between spacetime and $4$-momentum. 
However, the identification of $Spin(3,1)$ with
$Spin(1,3)$ in GR is inconsistent with the fact that the SMPP 
breaks $Spin(3,1)$ into $Spin(2)$ and  $Spin(1,1)$. In other words, if we maintain the local $Spin(1,3)$ symmetry of spacetime and $4$-momentum, 
then the Standard Model effectively breaks the global spacetime symmetry, by choosing a particular direction in space to define the mass gauge. Presumably this is the
direction (`down') of the gravitational field, which has always been used to define mass.

But now we can see immediately how the incompatibility of the SMPP with GR arises: the SMPP takes the splitting of $Spin(3,1)$ into $Spin(2)$ and $Spin(1,1)$ to be universal,
while GR takes it to be contingent on the local gravity. Moreover, there is a clear structure by which GR can generalise the SMPP to take account of a varying direction
of gravitational field. Indeed, it is clear from this analysis that the model predicts that the
SMPP does not work correctly if the direction of the gravitational field changes over the extent of an experiment.

For example, $Spin(2)$ is used in this model to describe the generation symmetry of neutrinos, and $Spin(1,1)$ is used to describe mass. In the original Standard Model,
these two groups were entirely separate, and the neutrinos assumed to be massless. However, experiments revealed that 
under certain circumstances, neutrinos could change generation within $Spin(2)$, which means they also change their relationship to $Spin(1,1)$. Hence it was deduced 
that neutrinos must have mass, since that is what $Spin(1,1)$ measures, and hence the mass was assumed to be the cause of the neutrino oscillations.

But in the context of GR, there is a completely different explanation. If the gravitational field is completely uniform, then there is no reason
to suppose that neutrinos will oscillate between generations. But
as soon as the gravitational field changes, which it always does in practice, the neutrinos will oscillate. The illusion of 
neutrino mass then arises from `measuring' the mass in a direction different from
the local direction `down'. Thus we are led to the conclusion that neutrino oscillations are in effect a general relativistic correction to the SMPP.

More generally, any particle experiment in which the separation of space into horizontal and vertical directions is not sufficiently uniform is liable to exhibit
an `anomaly' with respect to the SMPP. An experiment with diameter $6$ metres is sufficient to exhibit a 1ppm anomaly, that is the sine of the angle by which the
gravitational field changes. Examples of experiments which have reported anomalies of this magnitude include the following:
\begin{itemize}
\item Anomalous decay of long-eigenstate kaons into two pions instead of three \cite{CPexp}. Interpreted in SMPP as CP-violation, but 
according to GR caused by an inconsistency in the direction
in which the CP symmetry is applied, relative to the gravitational field. The experiment was approximately 17 metres across, and the anomaly was measured at 
between 2 and 3ppm. Other kaon anomalies \cite{kaonanomaly,Kaon2} may be due to more subtle differences between the North--South and East--West directions.
\item Anomalous magnetic moment of the muon \cite{muontheory,muonHVP,Fodor,muong-2}. 
The experiments have all apparently been $15$ metres across, and have measured an anomaly in muon $g-2$,
relative to the Standard Model prediction, of about
2.5 ppm.
\end{itemize}

In this model, all these anomalies arise from the difference between the Lorentz group in particle physics, which acts on only one Lorentz $4$-vector, and the Lorentz
group in relativity theory, which acts on two $4$-vectors. Therefore the model interprets the particle physics version as acting on $4$-momentum, measured
with respect to a static local spacetime, independent of gravity. The relativistic version, on the other hand, incorporates a dynamic spacetime 
in order to embed the local spacetime in a gravitational field. The solution to the incompatibility between GR and the SMPP therefore lies in \emph{either} converting
GR to work with a static spacetime \emph{or} converting the SMPP to work with a dynamic spacetime that takes the gravitational field into account. 
The former permits a quantisation of gravity in which the 
curvature of spacetime is replaced by a theory of neutrino oscillations, while the latter allows the curvature of spacetime to modify the parameters of the SMPP.
Both are valid interpretations of the $E_8$ model.

The notation has been chosen so that the particle physics version of the Lorentz group acts on the labels $U,IL,JL,KL$, while the general relativistic version acts
also on $L,I,J,K$. If we therefore break the symmetry using $D_{I,J}$ for the complex structure, then we have put the mass gauge into $D_{K,L}$ associated to the
`down' direction $K$. An alternative notation is to use $I+J+K$ for the `down' direction, so that $I$, $J$ and $K$ then become labels for the three generations
of neutrinos, and the differences $I-J$, $J-K$ and $K-I$ describe neutrino oscillations. But at the same time, these differences describe directions in the horizontal plane,
which only become apparent when the scale is large enough to notice that the `horizontal plane' is not flat.

In standard physics, the (Weak) Equivalence Principle is used to transfer the gravitational mass, measured in the vertical direction,
to the inertial mass, measured in the horizontal directions. This equivalence is enforced by GR, but is unfortunately inconsistent
with the SMPP, at least when embedded in $E_{8(-24)}$. That is, the equivalence can be maintained locally, provided there is sufficient neutrino oscillation
to ensure that there is no measurable correlation between neutrino generations and the direction of the gravitational field. But if gravity is sufficiently weak
that quantum effects become noticeable, and there is not enough neutrino oscillation to eliminate any inherent bias between the three generations,
then the Equivalence Principle can no longer be supported.

A further example in which the local equivalence of inertial and gravitational mass does not hold globally is in the evolution over time of the
gravitational environment of the Earth. Over the fifty years since the adoption of the SMPP, the gravitational environment has changed subtly by a small decrease
in the angle of tilt of the Earth's axis. This has a subtle effect on the relationship between the North/South, East/West and up/down axes, which means that the current
equivalence between inertial and gravitational mass is not quite the same as it was fifty years ago. Moreover, these changes in equivalence are not exactly the same
in all parts of the world. Any experiment that relies implicitly on an exact global equivalence, rather than a slowly varying local equivalence, is liable to
exhibit anomalies. Examples of experiments that do indeed exhibit such anomalies include the following.
\begin{itemize}
\item Inconsistent measurements of the Newtonian gravitational constant $G$ over time or between different locations \cite{Gillies,newG,QLi}.
Such inconsistencies, if real, can be interpreted in two ways. If we take the global equivalence of active gravitational mass and inertial mass as axiomatic,
then we must interpret $G$ as a variable rather than a constant. On the other hand, if we take the universality of $G$ as axiomatic, then we must interpret
the gravitational mass as a non-constant function of the inertial mass. Both interpretations are equally valid.
\item Unexplained drift in the measured masses of copies of the International Prototype Kilogram (IPK) in different parts of the world.
The relationship between the mass gauge and neutrino generations in the $E_{8(-24)}$ model suggests that it may be the mass of the electron that is causing this effect.
If the change in gravity is affecting the balance between the three neutrino generations, then it may also be affecting the balance between the
electron generations, causing the electron gravitational mass to drift apart from the electron inertial mass, leaving the proton and neutron masses
unaffected.
\end{itemize}

\section{The Pati--Salam model}
The splitting of $Spin(12,4)$ into $Spin(8)\otimes Spin(4,4)$, and the proposal to restrict from $Spin(4,4)$ to $GL_4(\RR)$ to implement GR in the model,
suggests restricting also from $Spin(8)$ to $SU(4)$. The triality symmetry of $Spin(8)$ rotates three copies of $SU(4)$, one of which acts on the $SO(8)$ vectors as
$SO(6)$, while the other two act on vectors as $SU(4)$. The usual assumption of the $Spin(10)$ GUT is that the Pati--Salam model embeds in it as
$Spin(6)\otimes Spin(4)$. 
The proposal here is different: we apply the triality symmetry to 
convert $SO(6)$ into $SU(4)$ as a subgroup of $SO(10)$ rather than $Spin(10)$.

In this way we get a group that acts on $SO(10)$ vectors as $SU(4)$, and therefore
acts on one half of the spinors as $SO(6)$, and on the other half as $SU(4)$. 
It is easy to restrict from the Georgi--Glashow $SU(5)$ to calculate that we have an action as $SO(2)\times SO(6)$ on neutrinos and up quarks,
with weak isospin $1/2$, and an action
as $U(4)$ on electrons and down quarks, with weak isospin $-1/2$. Hence we obtain a splitting of the spinors into $Y$ and $Z$ types, one of which has positive
weak isospin, and the other negative. 
In our notation, the scalars in $U(4)$ are generated by $E_l$, and therefore the neutrinos and up quarks are of $Y$ type, and the electrons and down quarks
are of $Z$ type.

Now the triality symmetry in $E_{8(-24)}$ acts simultaneously on both $Spin(8)$ and $Spin(4,4)$, not on each independently, so there is a corresponding
set of three subgroups of $SO(4,4)$, consisting of two (chiral) copies of $GL(4,\RR)$ and one copy of $SO(1,1)\times SO(3,3)$. 
Again, our choice of $GL(4,\RR)$ acts as $SO(1,1)\times SO(3,3)$ on one half of the spinors, either $Y$ type of $Z$ type.
In GR, the group $SO(3,3)$ acts on 
the field strength tensor, so we are led to conclude that the gravitational field is represented by massless (spin $1/2$) spinors rather than (spin $2$) tensors. 
Hence the field strength tensor must be represented by neutrinos, so is of $Y$ type.

This proposal therefore provides a physical mechanism for quantum gravity to work via exchange of neutrinos and antineutrinos, and provides a
mathematical context in which to do the requisite calculations. But it conflicts with two sacred cows: one is the assumption that forces must always be
mediated by bosons, not fermions, and the other is the Coleman--Mandula Theorem. Since the latter relies on the former, we only need to deal with that.
It suffices to point out that this is only an assumption, based on experience with electromagnetism (spin $1$ photons), the weak force
(spin $1$ intermediate vector bosons, the $Z$, $W^+$ and $W^-$ bosons) and the strong force (spin $1$ gluons). It would be reasonable to assume
that quantum gravity is also mediated by bosons, but no such bosons have been found experimentally, which at the very least casts some doubt on the 
validity of the assumption.

To see where the rest of the Pati--Salam gauge group lies, that is $Spin(4) = SU(2)_L\times SU(2)_R$, we must look in the centralizer of $SU(4)$. % which
By applying the triality automorphism we deduce that this centralizer is $Spin(6,4)$, of which only $Spin(4,4)$ preserves the distinction between fermions and bosons.
If we assume the Coleman--Mandula Theorem, then the gauge group commutes with $Spin(1,3)$,
and therefore lies in $Spin(3,1)$. But this has the wrong signature, since the gauge group is supposed to be compact $Spin(4)$.
There is no obvious solution to this problem, except to say that the full Pati--Salam gauge group does not embed in $E_{8(-24)}$. There is of course
a copy of $Spin(3)$ available for one of the $SU(2)$ factors, or some combination. 
But no such copy of $SU(2)$ acts on weak isospin, so the gauge group of the weak force surely cannot be here. 

Indeed, it is where it always was, including
the $W$ bosons in $X_I+X_{Jl}$ and $X_J-X_{Il}$, which swap the $Y$ and $Z$ spinors and therefore 
negate the weak isospin. To investigate further, let us combine this Georgi--Glashow version of weak $SU(2)$ with the (modified) Pati--Salam $Spin(3,1)$ into
the group $Spin(5,1)$ acting on the $6$ labels $u,l,I,J,K,L$. Since $Spin(5,1)\cong SL(2,\HH)$, we can write the generators as $2\times 2$ quaternion matrices,
and use the quaternion structure of $I,J,K$ in this description. For example, we can write the generators $D_l$, $X_I$, $X_J$, $X_K$, $X_L$ as
\begin{align}
\begin{pmatrix}0 & 1\cr -1 & 0\end{pmatrix},\quad
\begin{pmatrix}I & 0\cr 0 & -I\end{pmatrix},\quad
\begin{pmatrix}J & 0\cr 0 & -J\end{pmatrix},\quad
\begin{pmatrix}K & 0\cr 0 & -K\end{pmatrix},\quad
\begin{pmatrix}0 & 1\cr 1 & 0\end{pmatrix}
\end{align}
respectively.

Then the Georgi--Glashow version of $SU(2)_W$ is generated by
 \begin{align}
 X_{I}+X_{Jl} & =\begin{pmatrix}I&J\cr J&-I \end{pmatrix},\quad
 X_{J}-X_{Il}  =\begin{pmatrix}J&-I\cr -I&-J \end{pmatrix},\cr
   D_{I,J}-D_{l} & =\begin{pmatrix}K&-1\cr 1&K \end{pmatrix},
 \end{align}
 The standard way to add a `right-handed' counterpart is to change the signs of the off-diagonal entries,
 which gives altogether a group $SO(4)$ acting on the labels $u,l,I,J$.
 
The complex (modified Pati--Salam) version $Spin(3,1)$ is generated by 
\begin{align}
&D_{I,L} = \begin{pmatrix} 0 & I\cr -I & 0\end{pmatrix}, \quad
D_{J,L} = \begin{pmatrix} 0 & J\cr -J & 0\end{pmatrix}, \quad
D_{K,L} = \begin{pmatrix} 0 & K\cr -K & 0\end{pmatrix}, \cr
&D_{J,K} = \begin{pmatrix} I & 0 \cr 0 & I \end{pmatrix},\quad
D_{K,I} = \begin{pmatrix} J & 0 \cr 0 & J \end{pmatrix},\quad
D_{I,J} = \begin{pmatrix} K & 0 \cr 0 & K \end{pmatrix}
\end{align}
acting on the labels $I,J,K,L$.
There is another copy of $Spin(3,1)$ that may also be of use, swapping $I,J$ with $u,l$ to get an action on $u,l,K,L$
\begin{align}
D_{K,L}= \begin{pmatrix} 0 & K\cr -K & 0\end{pmatrix},  X_L=\begin{pmatrix}0 & 1\cr 1 & 0\end{pmatrix}, X_{lL}= \begin{pmatrix} 1 & 0\cr 0 & -1\end{pmatrix} ,\cr
D_l=\begin{pmatrix}0 & 1\cr -1 & 0\end{pmatrix}, X_K=\begin{pmatrix}K & 0\cr 0 & -K\end{pmatrix},X_{lK}= \begin{pmatrix} 0 & K\cr K & 0\end{pmatrix}, 
\end{align}
This version makes the symmetry-breaking of $I,J,K$, via restriction from $\HH$ to $\CC$, more obvious. 
In both cases, $L$ is used to complexify $Spin(3)=SU(2)$ to $Spin(3,1)=SL(2,\CC)$, and hence allow mass to be added. 

On the other hand, if we want an explicit implementation of the compact
 Pati--Salam $SU(2)_L\times SU(2)_R$,
 then we can combine the `left-handed' one above, generated by $D_{J,K}$, $D_{K,I}$ and $D_{I,J}$, with the `right-handed' one generated by
 $D_{JL,KL}$, $D_{KL,IL}$ and $D_{IL,JL}$. Alternatively, we can apply triality to this part of the Pati--Salam model as well, so that $D$ becomes $E$,
 in which case the group becomes the compact part of the group $GL_4(\RR)$ acting on the $8$ labels $U,I,J,K,L,IL,JL,KL$.
 In that case, the only reasonable interpretation of
this copy of $Spin(4)$ is that it has nothing to do with the weak force,
but everything to do with gravity. Hence we can perhaps attempt
to relate it to General Relativity.

\section{General relativity}
The field strength tensor is represented by the six components
\begin{align}
Y_I,Y_J,Y_K,Y_{IL},Y_{JL},Y_{KL}
\end{align}
independently of the $\OO$ labels that distinguish particle types.
The scalars that define $GL(4,\RR)$ are generated by $E_L$, so that the $15$ generators for $SL(4,\RR)$ are
\begin{align}
\label{SL4Rgens}
& D_I-D_{IL,L}, D_J-D_{JL,L}, D_{K}-D_{KL,L},\cr
& D_{J,K}-D_{JL,KL},D_{K,I}-D_{KL,IL},D_{I,J}-D_{IL,JL},\cr
& D_{IL}-D_{I,L}, D_{JL}-D_{J,L},D_{KL}-D_{K,L},\cr
& D_L-D_{I,IL}, D_L-D_{J,JL}, D_L-D_{K,KL},\cr
& D_{J,KL}+D_{K,JL}, D_{K,IL}+D_{I,KL},D_{I,JL}+D_{J,IL}.
\end{align}
The first two rows are the rotations, generating $SO(4)$. The second and third rows generate a copy of $SO(3,1)$, in which the terms in
$IL$, $JL$, and $KL$ generate the copy of $SL(2,\CC)$ that is used for the Lorentz group in the particle physics (Georgi--Glashow--Penrose) part of
our model.
The remaining terms in the second row represent neutrino oscillations, and show how these depend on the rotation of the gravitational field.
The remaining terms in the third row represent masses associated to the three generations (of electrons), and show how
General Relativity determines these masses.

The neutrinos themselves combine a generation label $I,J,K$ with a momentum label $IL,JL,KL$, and therefore create waves in the field-strength tensor,
and propagate the gravitational field. The Pati--Salam $SU(2)_L$ and $SU(2)_R$ act on these $3$-spaces, and can therefore be generated by
\begin{align}
&E_{I,J}, E_{J,K}, E_{K,I}\cr
& E_{IL,JL}, E_{JL,KL}, E_{KL,IL}.
\end{align}
There are linear equations \cite{WDM} that express these rotations in terms of the $D$ type elements given above. They are sums and differences of
the rotations in the first two rows. Hence, for example, a neutrino oscillation of the form $E_{I,J}$ mixes together the four components $D_{I,J}$, $D_K$, $D_{IL,JL}$
and $D_{KL,L}$. These rotations represent physical rotations in relativistic spacetime, and are therefore subject to tidal effects of (quantum) gravity.

The $X$ labels represent spacetime and/or $4$-momentum, either both locally, or one measured in two different frames of reference, depending on
the desired interpretation. In addition, there are $12$ more elements with $D$ labels, that form the part of $Spin(4,4)$ that lies outside $GL(4,\RR)$.
These are the elements obtained by changing the signs in the last four rows of (\ref{SL4Rgens}):
\begin{align}
& D_{J,K}+D_{JL,KL},D_{K,I}+D_{KL,IL},D_{I,J}+D_{IL,JL},\cr
& D_{IL}+D_{I,L}, D_{JL}+D_{J,L},D_{KL}+D_{K,L},\cr
& D_L+D_{I,IL}, D_L+D_{J,JL}, D_L+D_{K,KL},\cr
& D_{J,KL}-D_{K,JL}, D_{K,IL}-D_{I,KL},D_{I,JL}-D_{J,IL}.
\end{align}
As a representation of $SL(4,\RR)$ this splits up into two copies of the $6$-dimensional real representation, which can be distinguished as the
eigenspaces of the scalar, say $E_{I+J+K,L}$, but are isomorphic so can be mixed in arbitrary proportions.

The most important issue is to figure out where the mass gauge used in the SMPP comes from. That is, what do the labels $I,J,K$ actually signify?
They are quaternions, of course, so that the structure of the quaternions describes the generation structure of the electrons. A suitable coordinatisation
of the quaternions consistent with mass is given in \cite{finite}, and reproduced in Section~\ref{electron} above. From this, the actual mass gauge quaternion used
in the SMPP was calculated, gauged in units of eV/$c^2$:
\begin{align}
Q = -1293332 + 835200377I - 835982710J + 940347753K
\end{align}
The real part here is the difference between the proton and neutron masses.
Normalizing to norm $3$ is achieved by dividing by $871911186$ eV to give a dimensionless mass gauge quaternion
\begin{align}
Q_0 = -.00148333 + .95789616I -.95879342J + 1.07849030K
\end{align}
It is noticeable that this quaternion is close to $I-J+K$, but with a small difference attributable mainly to the mass of the muon.
 
 The electrons all decay quickly to the first generation state, that is the lowest mass state, i.e. the lowest gravitational energy state. This state is
 perpendicular to $K$, so that $K$ is labelling the gravitational field, or the directions up and down. Similarly, $I$ and $J$ denote the directions North/South
 and East/West, in some order. Hence the coefficients of $I$ and $J$ denote horizontal masses, with a small difference of about $.01\%$ between them.
 The mass used in general physics, then, is a compromise between the $I$, $-J$ and $K$ coefficients, that enforces the equivalence of
 horizontal (inertial) and vertical (gravitational) mass. In normal bulk matter, there are no muons to worry about, so that the difference between horizontal and
 vertical masses disappears. But the much smaller difference between North/South and East/West masses is still there---it is the difference between
 the neutron mass and the sum of the proton and electron masses.
 
 It follows from this analysis that the mass ratios of electron, proton and neutron can in principle be derived from GR, by computing in detail
 the effects of a varying gravitational field on the variables use in the SMPP. The difference between the proton and neutron masses appears in the
 real part of the mass gauge quaternion, and is therefore not affected by changes in spatial directions, but only by changes in time. The transformations in
 $SL(4,\RR)$ convert between a local time on Earth and a global time in the larger Solar System. The former is measured in days, the latter in years, so that
 the appropriate dimensionless parameter that goes into the calculation is the number of days in a year, say $365.24$. To a first approximation, the neutron/proton
 mass ratio is calculated as
 \begin{align}
 m(n)/m(p) &\approx 1 + 1/(2\times 365.24)\cr
 &\approx 1.001369
 \end{align}
 compared to the experimental value $1.001378$.
 
 As for the electron mass, this does depend on a difference in direction between $K$ on one hand, and $I$ and $J$ on the other.
 Again it is a direction that relates the local day to the global year, and is therefore the tilt of the Earth's axis, of around $23.437^\circ$. To a first approximation, the calculation
 gives
 \begin{align}
 m(p)/m(e) &\approx 2\times 365.24/\sin(23.437^\circ)\cr
 & \approx 1836.57
 \end{align}

This  compares to the experimental value $1836.15$, with a difference of around $.02\%$. 
However, it is worth noting that the angle of tilt varies significantly, and the angle that is required to get the
 mass ratio exactly correct is around $24.443^\circ$. This is roughly the angle of tilt in the early 1970s when the SMPP was being developed.
 It is therefore this larger angle that goes into the calculations to get the mass values that are used in the SMPP. Since the SMPP has a fixed mass gauge,
 not a variable one, the calculations are not updated to take account of the new gravitational mass value for the electron. All such changes in the relationship
 between gravitational $K$ and inertial $I,J$ are subsumed instead into mixing angles using $Spin(2)$ generated by $D_{I,J}$.
 
 This analysis shows that the Weak Equivalence Principle (WEP) that equates inertial mass with active gravitational mass is a local equivalence that must be calibrated locally.
 This can be converted into a global equivalence for \emph{one} type of particle only. In this model, I have taken the neutron to do this calibration.
 In standard physics, it is the carbon-12 atom that is chosen. But in practice, a platinum-iridium alloy was used to calibrate (or define) mass for $130$ years,
 leading to some unexplained anomalies. The SMPP works only with inertial mass, and has measured (or defined) these very accurately. But the SMPP
 is inconsistent with the WEP, which is the basis for GR, so that active gravitational masses do not precisely match standard inertial masses.
 Nevertheless, the standard inertial masses are ultimately based on terrestrial gravitational masses, 
 using a `standard' gravitational field,
 which closely matches the actual field in around 1973.

\section{Mediators}
In the SMPP the mediators for the forces lie in the adjoint representation of the gauge group. However, the assumption that they always lie in the adjoint representation
poses a problem for GUTs, as was recognised from the outset. If mediators for the Georgi--Glashow SU(5) model lie in the adjoint representation, then there are $12$
extra mediators that have never been detected, and there are forces that cause protons to decay, a phenomenon that likewise has never been detected.
The problem is the same in $E_8$ models, in which the unified `gauge group' can be either $Spin(11,1)$, as here, or $Spin(9,3)$, as in \cite{octions}. We do not want to
increase the count of mediators from the $12$ in the SMPP to $66$, so the only option is to follow the example of \cite{octions}, and index the mediators with the
$12$-dimensional vector representation of the appropriate real form of $SO(12)$. The alternative is to reject the whole idea of a GUT, or any other kind of unification.

This proposed change makes no theoretical difference to gauge groups $U(1)$ and $SU(2)$, in which the vector representation can be identified with the adjoint representation, but
it makes a difference to $SU(3)$, replacing the $8$-dimensional real adjoint by a $6$-dimensional real vector. In \cite{octions} this problem is left open, and only
six gluons are explicitly identified in the model. However, it can be seen that $E_8$ has room for all $12$ mediators, so with a slight adjustment to the interpretations
in \cite{octions}, it should be possible to include the remaining two gluons.

These two (colourless) gluons must however lie in scalar representations of $SU(3)$. One possibility, therefore, is to regard the colourless gluons,
that glue two quarks into mesons, as fundamentally different from the
coloured gluons, that glue three quarks into baryons. Another possibility is to use $SU(4)$ to unify the $8$ gluons into an irreducible representation.
 
 Either way, once we have used $8$ coordinates for the gluons, we are left with either $SO(1,3)$ or $SO(3,1)$ for the four electro-weak mediators. There is a clear
 separation between one photon and three intermediate vector bosons, but there is no clear choice for basis vectors, and therefore there is scope for describing
 a `mixing' between the two forces. The choice between the two possible signatures is really a choice as to which symmetries one wants to make explicit.
 The rationale of this paper has been to make the generation symmetry explicit, which led to the group $Spin(11,1)$. The rationale of \cite{octions} was to give more 
 prominence to the forces and mediators, which led to a preference for $Spin(9,3)$.
 
 The question then arises, how to use the signature $(9,3)$, and how to split the $12$ coordinates into forces and mediators.
 The choices made in \cite{octions} led to using $SL(3,\RR)$ instead of $SU(3)$ for the gauge group of the strong force, which is not only inconsistent with the SMPP,
 but also leads to problems identifying the colourless gluons. Therefore I propose instead to use the splitting $9+3$ as a splitting into $9$ massless and $3$
 massive bosons, and further split $9$ into $8$ gluons and one photon, so that the group splits into $Spin(8)\otimes Spin(1,3)$. Restriction from
 $Spin(8)$ to $SU(4)$ or $SU(3)$ can then be used for the strong force, and $Spin(1,3)=SL(2,\CC)$ is the complex form of $SU(2)$, which is used in the SMPP
 to give masses to the $W$ and $Z$ bosons, and potentially also describes electro-weak unification.
 
 This leads to using the labels $u,l$ for colourless gluons, and $i,il,j,jl,k,kl$ for coloured gluons, $U$ for the photon and $IL,JL,KL$ for the $W$ and $Z$ bosons,
 leaving $I,J,K,L$ for the Lorentz group to act on. The symmetry-breaking between the $Z$ and $W$ bosons then relates again to a standard definition of `vertical' and
 `horizontal' directions related to the gravitational field. This symmetry-breaking arises from the fact that in GR the Lorentz group acts on both vectors $(I,J,K,L)$ and
 $(U,IL,JL,KL)$, not just one of them. The precise way in which these two Lorentz vectors are related to each other is a problem for quantum gravity to resolve, but the
 fact that they are related follows just from the assumption that there is a unified model of gravity and particle physics in $E_8$.
 
 In particle physics, of course, one of $Spin(1,3)$ and $Spin(3,1)$ must be interpreted as the `left-handed' gauge group of the weak force, acting on weak isospin,
 while the other is interpreted as the Lorentz group, acting on spin. It is therefore tempting to describe the Lorentz group, and by extension spacetime itself, as
 `right-handed', as Woit has done \cite{WoitRH}, although it is not at all clear how to interpret this statement. 
 Presumably it refers to the nature of spacetime in the
 laboratory where particle experiments are performed, rather than to spacetime in the abstract or in general. 
 It may therefore be a local rather than global property of spacetime.
 
 Since the local shape of spacetime depends on the
 local properties of gravity and/or acceleration, the relevant accelerations of the laboratory 
 are the acceleration towards the Earth's axis due to the rotation of the Earth, the acceleration
 towards the Sun due to the Earth's orbit, and the acceleration towards the Moon due to the gravity of the Moon. The `right-handedness' must 
 presumably therefore refer to the handedness
 of this system of three rotations that defines our local spacetime, rather than to an abstract `chirality' that applies to the entire universe at once.
 If so, then there is an entirely different interpretation of the Wu experiment \cite{Wu} from the standard one, namely that observed chirality
 was in fact an observation of quantum gravity in action.
 
\section{Conclusion}
In this paper I have taken the finite symmetry groups defined by the scalars in the gauge groups $SU(3)$ and $SU(2)$ of the strong and weak interactions, 
plus the Lorentz group $SL(2,\CC)$, and
added to them a finite symmetry group $Z_3$ to describe the three generations of elementary fermions. 
I have shown that there is a unique embedding of these finite symmetries
in $E_{8(-24)}$, consistent with their fundamentally discrete character, and consistent with both gauge invariance and Lorentz covariance, and the
Coleman--Mandula Theorem \cite{ColemanMandula}. I have also shown
 that the centralizer of the finite symmetry group is essentially the Standard Model of Particle Physics, extended by an extra copy of $U(1)$ plus a
copy of $Spin(1,1)$. 

The generation symmetry then corresponds to a scalar of order $9$ in $SU(7,2)$, which introduces multiples of $40^\circ$ into the model, 
to add to the multiples of $90^\circ$ which arise from the factor of $i$ in the mass term in the Dirac equation.
These angles
can be used to relate the quark-mixing angles in the CKM matrix to the lepton-mixing angles in the PMNS matrix, and apparently reduce the
number of independent parameters in the SMPP. Of particular note is the surprising conclusion that the Cabibbo angle can be \emph{calculated} as $13.0240^\circ$ 
from the mass ratios of the electron, proton, neutron and muon only, without any other experimental data.

There is
also some indication that $E_8$ itself provides some predictive power, 
in addition to that which comes from various subgroups of $SU(7,2)$.
By separating a `matter' group $Spin(8)$ from a `spacetime/radiation' group $Spin(4,4)$ (including neutrinos)
we see an explicit distinction between a static Minkowski spacetime used
by the SMPP, and a dynamic Minkowski spacetime used by GR. This leads to wholesale predictions of anomalies arising from
not applying the necessary general relativistic corrections to the SMPP. 

It is hoped that it may eventually be possible to use such general relativistic
corrections to explain some of the otherwise unexplained mass parameters, in addition to the mixing angles already considered.
The key to successfully accomplishing this goal may be to appreciate that although the SMPP can successfully edit out any change in the
overall magnitude and direction of gravity, it is much harder to edit out the tidal effects of the quantum gravity of the Sun and the Moon.
Since these tidal effects necessarily have an impact on the mixing angles in the PMNS matrix, it is to be expected that they will also
have an impact on the masses that go into the calculation of the mixing angles \cite{universality}.

\end{document}